\documentclass[USenglish,twocolumn]{article}

\usepackage[russian,english]{babel}
\usepackage[big]{dgruyter}
\usepackage{upgreek}
\usepackage{amsmath,amsfonts,amssymb,pxfonts,eulervm,xspace}
\usepackage{graphicx}
\usepackage[title]{appendix}
\usepackage{natbib}
\usepackage{color}

\def\jref@jnl#1{{\rm#1\/}}
\def\actaa{\jref@jnl{Acta Astronomica}}
\def\aap{\jref@jnl{A\&A}}
\def\aapr{\jref@jnl{The Astronomy and Astrophysics Review}}
\def\aaps{\jref@jnl{Astronomy and Astrophysics Supplement Series}}
\def\aj{\jref@jnl{AJ}}
\def\apj{\jref@jnl{ApJ}}
\def\apjl{\jref@jnl{ApJL}}
\def\apjs{\jref@jnl{ApJS}}
\def\apss{\jref@jnl{Astrophysics and Space Science}}
\def\ao{\jref@jnl{Applied Optics}}
\def\araa{\jref@jnl{ARA\&A}}
\def\bain{\jref@jnl{BAN}}
\def\caa{\jref@jnl{Chinese Astronomy and Astrophysics}}
\def\cjaa{\jref@jnl{Chinese Journal of Astronomy and Astrophysics}}
\def\gca{\jref@jnl{Geochimica et Cosmochimica Acta}}
\def\jcp{\jref@jnl{Journal of Chemical Physics}}
\def\jqsrt{\jref@jnl{Journal of Quantitative Spectroscopy and Radiative Transfer}}
\def\mnras{\jref@jnl{MNRAS}}
\def\memras{\jref@jnl{Memoirs of the Royal Astronomical Society}}
\def\memsai{\jref@jnl{Memorie della Societa Astronomica Italiana}}
\def\na{\jref@jnl{New Astronomy}}
\def\nar{\jref@jnl{New Astronomy Reviews}}
\def\nat{\jref@jnl{Nature}}
\def\pasa{\jref@jnl{Publications of the Astronomical Society of Australia}}
\def\planss{\jref@jnl{Planetary and Space Science}}
\def\pasj{\jref@jnl{Publications of the Astronomical Society of Japan}}
\def\pasp{\jref@jnl{PASP}}
\def\physrep{\jref@jnl{Physics Reports}}
\def\pra{\jref@jnl{Physical Review A}}
\def\prd{\jref@jnl{Physical Review D}}
\def\pre{\jref@jnl{Physical Review E}}
\def\physrep{\jref@jnl{Physics Reports}}
\def\physscr{\jref@jnl{Physica Scripta}}
\def\qjras{\jref@jnl{Quarterly Journal of the Royal Astronomical Society}}
\def\rmxaa{\jref@jnl{Revista Mexicana de Astronomia y Astrofisica}}
\def\skytel{\jref@jnl{Sky and Telescope}}
\def\solphys{\jref@jnl{Solar Physics}}
\def\sovast{\jref@jnl{Soviet Astronomy}}
\def\ssr{\jref@jnl{Space Science Reviews}}
\def\zap{\jref@jnl{Zeitschrift fuer Astrophysik}}
\def\azh{\jref@jnl{Astronomicheskij Zhurnal}}
\def\procspie{\jref@jnl{Proc. SPIE}}

\def\akn{}

\newcommand{\MC}{\multicolumn}

\begin{document}

\articletype{Research Article{\hfill}Open Access}

\author*[1]{Oleg Malkov}       

\author[2]{Alexey Kniazev}    




\title{\huge Wide binary stars with non-coeval components}

\runningtitle{Wide binaries with non-coeval components}


\begin{abstract}
{We have estimated masses of components of visual binaries from
their spectral classification. We have selected pairs, where
the less massive component looks more evolved. Spectral observations
of some of such pairs were made, and at least one pair, HD~156331, was confirmed
to have components of different age. Since mass exchange is excluded
in wide binaries, it means that HD~156331 can be formed by the capture.}
\end{abstract}

\keywords{Binary Stars, Stellar Mass}

\journalname{Open Astronomy}
\DOI{DOI}
  \startpage{1}
  \received{..}
  \revised{..}
  \accepted{..}

  \journalyear{2017}
  \journalvolume{1}

\maketitle


{ \let\thempfn\relax
\footnotetext{\hspace{-1ex}{\Authfont\small
\textbf{Corresponding Author: Oleg Malkov:}} {\Affilfont Institute of Astronomy, Moscow, Russia; Email: malkov@inasan.ru}}
}

{ \let\thempfn\relax
\footnotetext{\hspace{-1ex}{\Authfont\small \textbf{Alexei Kniazev:    }} {\Affilfont South African Astronomical Observatory, Cape Town, South Africa; Sternberg Astronomical Institute, Moscow, Russia}}
}

\section{Introduction}

The formation of binary stars basically follows two scenarios: fission of rotating
molecular gas clouds during gravitational collapse, and inelastic collisions
of stars during the formation of young star clusters~\citep{2020PhyU...63..209T}.
A capture of a component from the field stars is not ruled out in principle either,
although it should be relatively rare.
Capture occurs when two stars pass close to each other
in the presence of a scattering medium that can take in excess kinetic energy,
to leave the two stars bound. This medium could be a third star,
a circumstellar disk, or the stars themselves, if the collision is close enough
to cause the tides to rise and fall.
Capture in the presence of a third body and ``tidal capture''
requires a high stellar density, which is atypical for field stars.

Capture in the presence of a stellar disk may play some role in the formation of wide systems,
because the capture cross section has to be on the order of the size of the disk
and thus lead to the formation of systems with large semi-axes
about 100 AU. \cite{1991MNRAS.249..584C} considered
the possibility of a large,
massive protostellar accretion disc playing a 
role in the formation of binary stars by enabling the capture of a passing star within a 
dense star-forming region. It was found that capture rates are too low to play a major role 
in all known star-forming environments, particularly when the probability of prior 
disc dispersal by the more frequent high-velocity interactions is taken into account. 

An indicator of the capture
could be the difference in the ages of the components. It is evident, in particular,
that in evolutionary wide systems (i.e., systems with no matter transfer
between components today or in the past) with components of the same
age, a less massive component cannot appear to be more evolved.

Our previous attempt to find pairs with non-coeval components was based on estimating
the durations of preMS and MS stages for stars of different masses.
In particular, we looked for wide pairs where a very low-mass secondary component
(with mass $m_2$ and duration of preMS stage $\tau_{\rm preMS}(m_2)$) is {\it already} an MS star,
and massive primary (with mass $m_1$ and duration of preMS + MS stages $\tau_{\rm MS}(m_1)$) is {\it still} an MS star.
We have found three candidates with $\tau_{\rm preMS}(m_2) > \tau_{\rm MS}(m_1)$, the results can be found
in~\cite{2000IAUS..200P.170M}.

The aim of the present study is to use another approach to find non-coeval pairs among
visual binaries. For indication of non-coevality we compared spectral classes
and masses of the components, estimated from the spectral classification.

{\akn
The structure of this paper is as follows: 
in Section~\ref{txt:sample} we describe our sample selection,
in Section~\ref{txt:spec_obs} we describe our observations for the first three objects and 
spectral data reductions.
Data analysis is described in Section~\ref{txt:analysis}, and
the results are discussed in Section~\ref{txt:results}.
Section~\ref{txt:summary} summarizes this paper.

\section{Sample selection}
\label{txt:sample}

The general method of this work is an applying the simple idea 
to find non-coeval pairs among visual binaries. 
For indication of non-coevality we compared spectral classes
and masses of the components, estimated from the spectral classification.
Applying this idea to the Sixth Catalog of Orbits of Visual Binary Stars, 
ORB6~\citep{2001AJ....122.3472H},
we found thirteen systems where less massive component looks more evolved,
and, consequently, the components are probably non-coeval~\citep{2020INASR...5..341M}.

We have made a search for additional data on these thirteen systems in
ORB6~\citep{2001AJ....122.3472H},
Catalogue of Stellar Spectral Classifications~\citep{2014yCat....102023S},
Multiple star catalogue, MSC~\citep{2018ApJS..235....6T},
as well as in the SIMBAD database.
The parameters of three systems presented in this paper are shown in Table~\ref{tab:pairs}.

\begin{table}
    \akn
    \caption{Systems under study. Pairs}
    \label{tab:pairs}
    \begin{center}
        \begin{tabular}{r|r|r|c}
            \hline
            Name      &   V, mag  &   $\varpi$, mas &   $\sigma_\varpi$ \\
            \hline                       
            HD 101379 &   5.095   &     8.397       &    0.507          \\
            HD 156331 &   6.267   &    16.703       &    0.048          \\
            HD 160928 &   5.871   &    13.053       &    0.599          \\
            \hline                       
        \end{tabular}
    \end{center}
Parallax $\varpi$ and visual brightness $V$ are taken from Gaia EDR3 and SIMBAD, respectively.
\end{table}

\section{Observations and Data Reduction}
\label{txt:spec_obs}

Three of these thirteen systems have been observed with the
Southern African Large Telescope \citep[SALT;][]{2006SPIE.6267E..0ZB,2006MNRAS.372..151O}
using the High Resolution Spectrograph 
\citep[HRS;][]{2008SPIE.7014E..0KB,2010SPIE.7735E..4FB,2012SPIE.8446E..0AB,2014SPIE.9147E..6TC}. 
The HRS is a thermostabilized double-beam echelle spectrograph, the entire optical part
which is housed in a vacuum to reduce the influence of temperature variations 
and mechanical interference.
The blue arm of the spectrograph covers the spectral range 3735--5580~\AA,
and the red arm covers the spectral range of 5415--8870~\AA, respectively.
The spectrograph is equipped with two fibers (object and sky fibers)
and can be used in the low (LR, R$\approx$14,000-15,000)
medium (MR, R$\approx$40,000--43,000) and high (HR, R$\approx$67,000--74,000) resolution modes.
For our observation HRS was used in MR, where for both the object and sky
fibers have 2.23~arcsec in diameter. 
The both blue and red arms CCD were read out by a single amplifier with a 1$\times$1 binning.
All additional details of observations are summarized in the Table~\ref{tab:Obs_log}.
Generally, each star was observed once, but in case HD\,101379 three spectra were obtained,
where for both HD\,156331 and HD\,160928 only one spectrum was obtained.
Exposures were selected in the way to accumulate Signal-to-Noise Ratio (SNR) more than 150
in the spectral region 4300--8800~\AA. Unfortunately, sensitivity of HRS drops down fast bluer of
4300~\AA\ and the final SNR in this spectral region is very hard to predict.

\begin{table}
    \akn
    \caption{Observational log}
    \label{tab:Obs_log}
    \begin{center}
        \begin{tabular}{r|c|c|c|c}
            \hline
      \MC{1}{c|}{Name}&   Date        &   Exposure     &   Seeing   &   SNR     \\
                      &               &\MC{1}{c|}{(s)} &   (arcsec) &           \\
            \hline                                                            
            HD\,101379 &  2021 July 14 & 3$\times$25   &    1.5     &  150--400 \\
            HD\,156331 &  2021 May  10 & 1$\times$40   &    1.1     &  150--280 \\
            HD\,160928 &  2021 May  10 & 1$\times$40   &    1.2     &  200--300 \\
            \hline                       
        \end{tabular}
    \end{center}
\end{table}

Three spectral flats and one spectrum of ThAr lamp
were obtained in this mode during a weekly set of HRS calibrations
that enough to get average external accuracy of 300 m~s$^{-1}$.
Method of analysis, described in Section~\ref{txt:analysis} needs 
to use spectra corrected for sensitivity curve. 
For this reason spectra of spectrophotometric standard from the list
of \citet{2017SALT.REPORT7}\footnote{\url{https://astronomers.salt.ac.za/software/hrs-pipeline/}}
were observed and used during HRS data reduction.

Primary reduction of the HRS data, including overscan correction,
bias subtractions and gain correction, was done with the SALT
science pipeline \citep{2010SPIE.7737E..25C}.
Spectroscopic reduction of the HRS data was carried out using
standard HRS pipeline and our own additions to it
described in details in \citep{2019AstBu..74..208K}.

\section{Spectral Data Analysis}
\label{txt:analysis}

\begin{table*}
    \akn
    \caption{Stellar parameters found with {\tt fbs} software}
    \label{tab:stars}
     \begin{tabular}{lrcrrcrrcl} \hline
System        & T$_{\rm eff}$ & $\log g$      & $\rm V \sin i$  & V$_{\rm hel}$  & Weight in        & $M_V$           & $\rm [Fe/H]$     & $\rm E(B-V)$  & St. lib.    \\
              &   (K)         & (cm~s$^{-1}$) & (km~s$^{-1}$)   & (km~s$^{-1}$)  & V band           & (mag)           &   (dex)          &   (mag)       &             \\
\hline                                                                                                                                                              
HD\,101379 A  &~5160$\pm$100  & 4.1$\pm$0.19  &  3.0$\pm$0.2    &--6.1$\pm$0.3   & 0.73$\pm$0.01    & -0.69$\pm$0.13  &    0.42$\pm$0.18 & 0.24$\pm$0.01 & Coelho      \\
HD\,101379 A  &~5000$\pm$25~  & 3.8$\pm$0.01  & 10.4$\pm$0.2    &--5.9$\pm$0.2   & 0.72$\pm$0.01    & -0.70$\pm$0.13  &    0.19$\pm$0.01 & 0.25$\pm$0.01 & Phoenix     \\
\hline                                                                                                                
HD\,101379 B  &10460$\pm$55~  & 4.6$\pm$0.08  &104.5$\pm$2.3    & 30.5$\pm$0.4   & 0.27$\pm$0.01    &  0.39$\pm$0.13  &    0.42$\pm$0.18 & 0.24$\pm$0.01 & Coelho      \\ 
HD\,101379 B  &10230$\pm$180  & 3.9$\pm$0.03  & 90.3$\pm$0.9    & 30.3$\pm$0.3   & 0.28$\pm$0.01    &  0.32$\pm$0.13  &    0.19$\pm$0.01 & 0.25$\pm$0.01 & Phoenix     \\ 
\hline\hline                                                                                                          
HD\,156331 A  &~6000$\pm$10~  & 4.0$\pm$0.01  & 35.1$\pm$0.2    &  9.8$\pm$0.2   & 0.72$\pm$0.01    &  2.74$\pm$0.01  &$-$0.28$\pm$0.01  & 0.00$\pm$0.01 & Coelho      \\
HD\,156331 A  &~5990$\pm$10~  & 3.8$\pm$0.06  & 36.8$\pm$0.2    &  9.9$\pm$0.1   & 0.68$\pm$0.01    &  2.80$\pm$0.01  &$-$0.20$\pm$0.02  & 0.00$\pm$0.01 & Phoenix     \\
\hline                                                                                                                                                             
HD\,156331 B  &~8190$\pm$20~  & 3.8$\pm$0.03  & 60.8$\pm$0.3    & 33.7$\pm$0.2   & 0.28$\pm$0.01    &  3.76$\pm$0.01  &$-$0.28$\pm$0.01  & 0.00$\pm$0.01 & Coelho      \\
HD\,156331 B  &~8010$\pm$20~  & 4.0$\pm$0.04  & 60.0$\pm$0.3    & 31.1$\pm$0.3   & 0.32$\pm$0.01    &  3.62$\pm$0.01  &$-$0.20$\pm$0.02  & 0.00$\pm$0.01 & Phoenix     \\ 
\hline\hline                                                                                                                                                    
HD\,160928 A  &~8270$\pm$10~  & 3.7$\pm$0.02  & 238.4$\pm$0.7   & $-$7.3$\pm$0.3 & 0.79$\pm$0.01    &  1.71$\pm$0.10  & $-$0.34$\pm$0.01 & 0.00$\pm$0.01 & Coelho      \\
HD\,160928 A  &~8450$\pm$20~  & 3.9$\pm$0.01  & 225.0$\pm$0.8   & $-$8.5$\pm$1.3 & 0.73$\pm$0.01    &  1.79$\pm$0.10  & $-$0.15$\pm$0.03 & 0.00$\pm$0.02 & Phoenix     \\
\hline                                                                                                                                                             
HD\,160928 B  &~6400$\pm$20~  & 4.5$\pm$0.06  & 173.5$\pm$0.5   &$-$15.4$\pm$0.6 & 0.21$\pm$0.01    &  3.14$\pm$0.10  & $-$0.34$\pm$0.01 & 0.00$\pm$0.01 & Coelho      \\
HD\,160928 B  &~6880$\pm$60~  & 4.8$\pm$0.01  & 190.3$\pm$1.5   &$-$16.5$\pm$1.6 & 0.27$\pm$0.01    &  2.87$\pm$0.10  & $-$0.15$\pm$0.03 & 0.00$\pm$0.02 & Phoenix     \\ 
\hline\hline                                                                                                                                                     
%
  Errors     &   300.         & 0.35          & 10.             &    1.3         &       0.02       &                 &   0.06            &        0.01   &            \\
\hline
    \end{tabular}
\end{table*}

\begin{figure}[]
    \includegraphics[width=7.7cm]{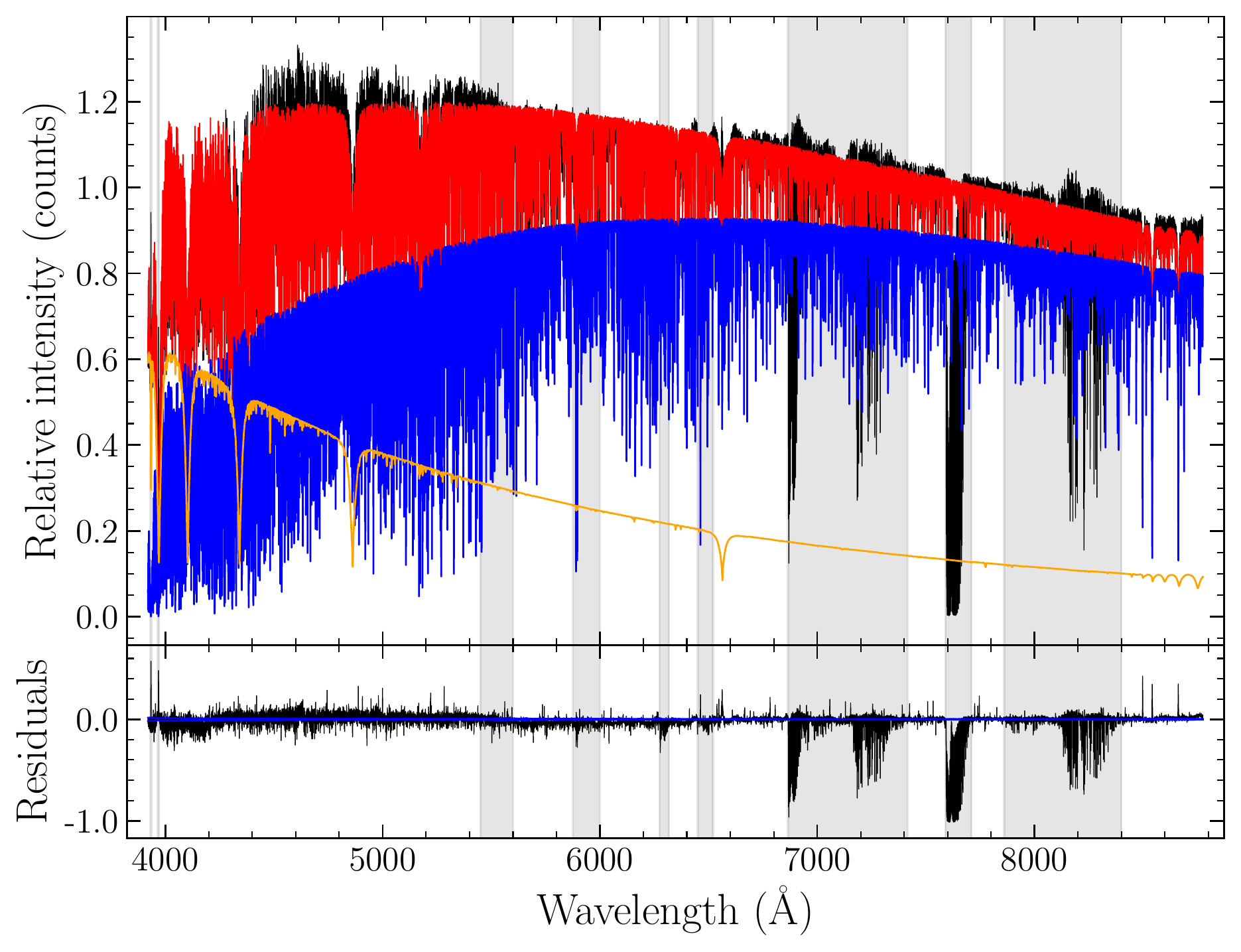}
    \includegraphics[width=7.7cm]{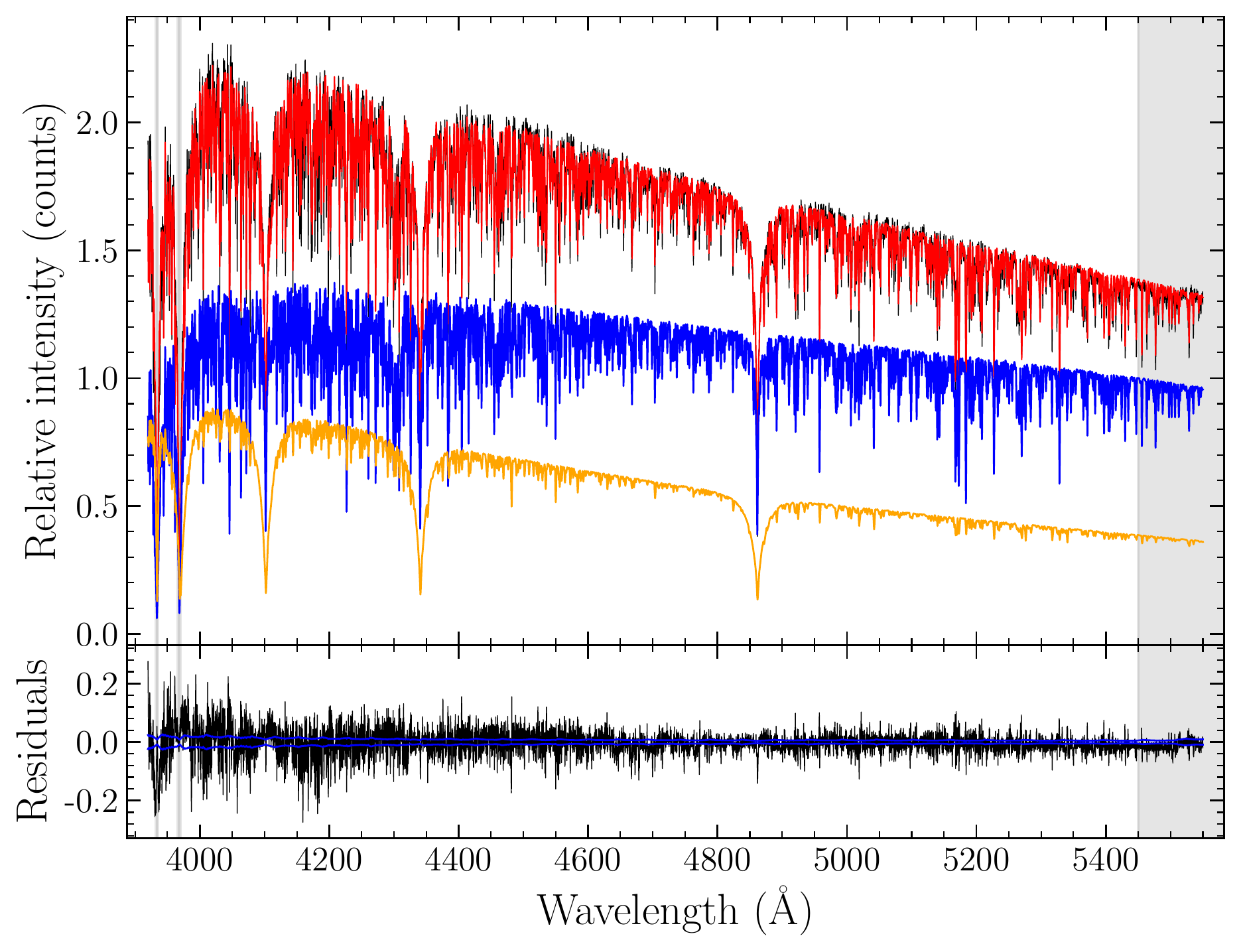}
    \includegraphics[width=7.7cm]{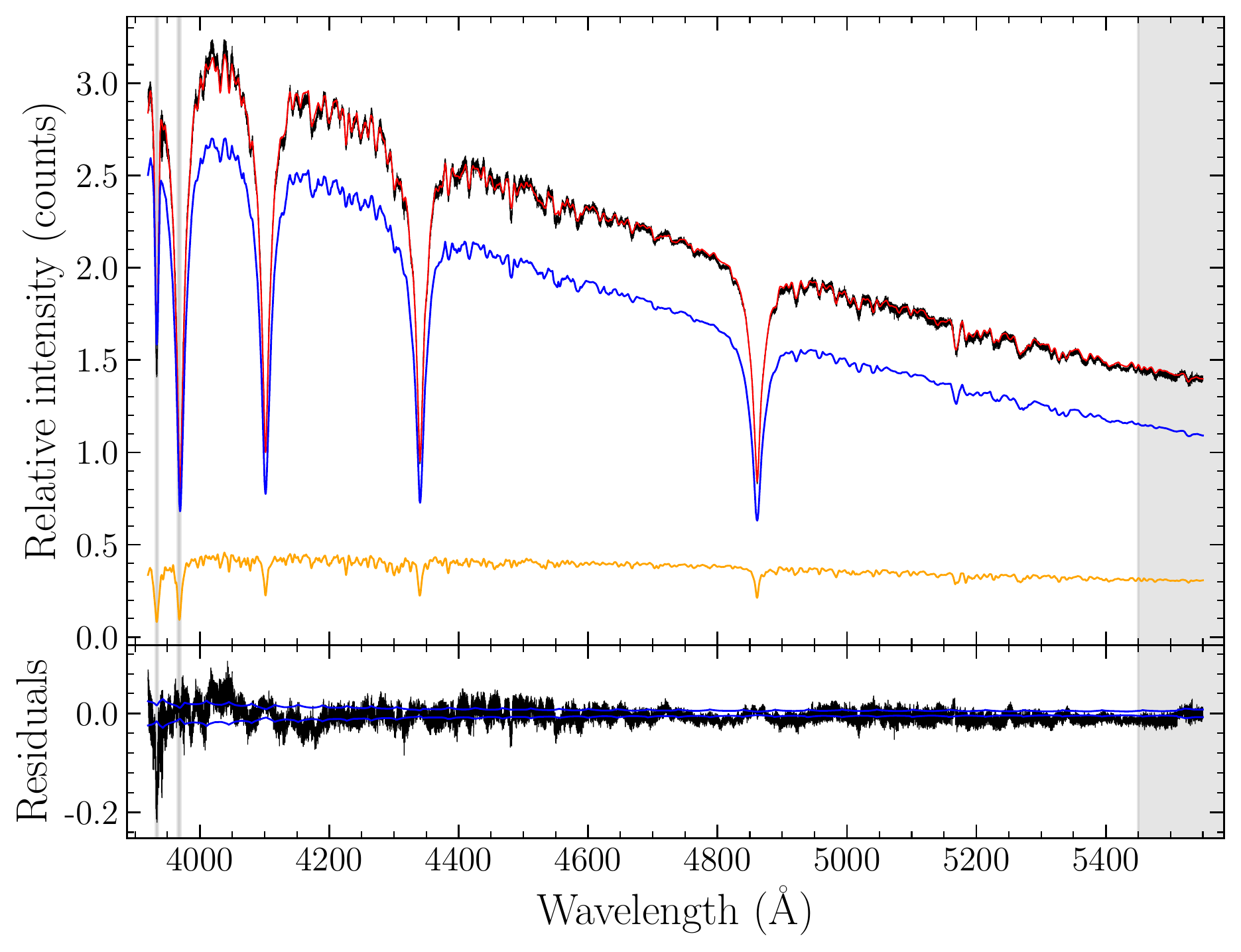}
    \caption{\akn The results of the fbs fit all stars from this work: HD\,101379 (top), 
             HD\,156331 (middle) and HD\,160928 (bottom).
             Each panel consists of two sub-panels: {\bf the top one}
        shows the result of the fit in the spectral region 3900-8870~\AA.
        The observed spectrum is shown in black.
        Both found components are shown in blue and orange respectively, 
        where their sum is shown in red.
        {\bf the bottom one} shows the difference between the observed spectrum and its model in black,
        altogether with errors that were propagated from the HRS data reduction (continuous dark blue line).
        Grey vertical areas mark the spectral ranges that were excluded from the fit for different reasons,
        mainly due to bands of lines from the Earth atmosphere. Only blue spectra are shown for 
        HD\,156331 and HD\,160928 just for more details.
    }
    \label{fig:spectrum}
\end{figure}
}

\begin{figure}[]
    \centering
    \includegraphics[width=8cm]{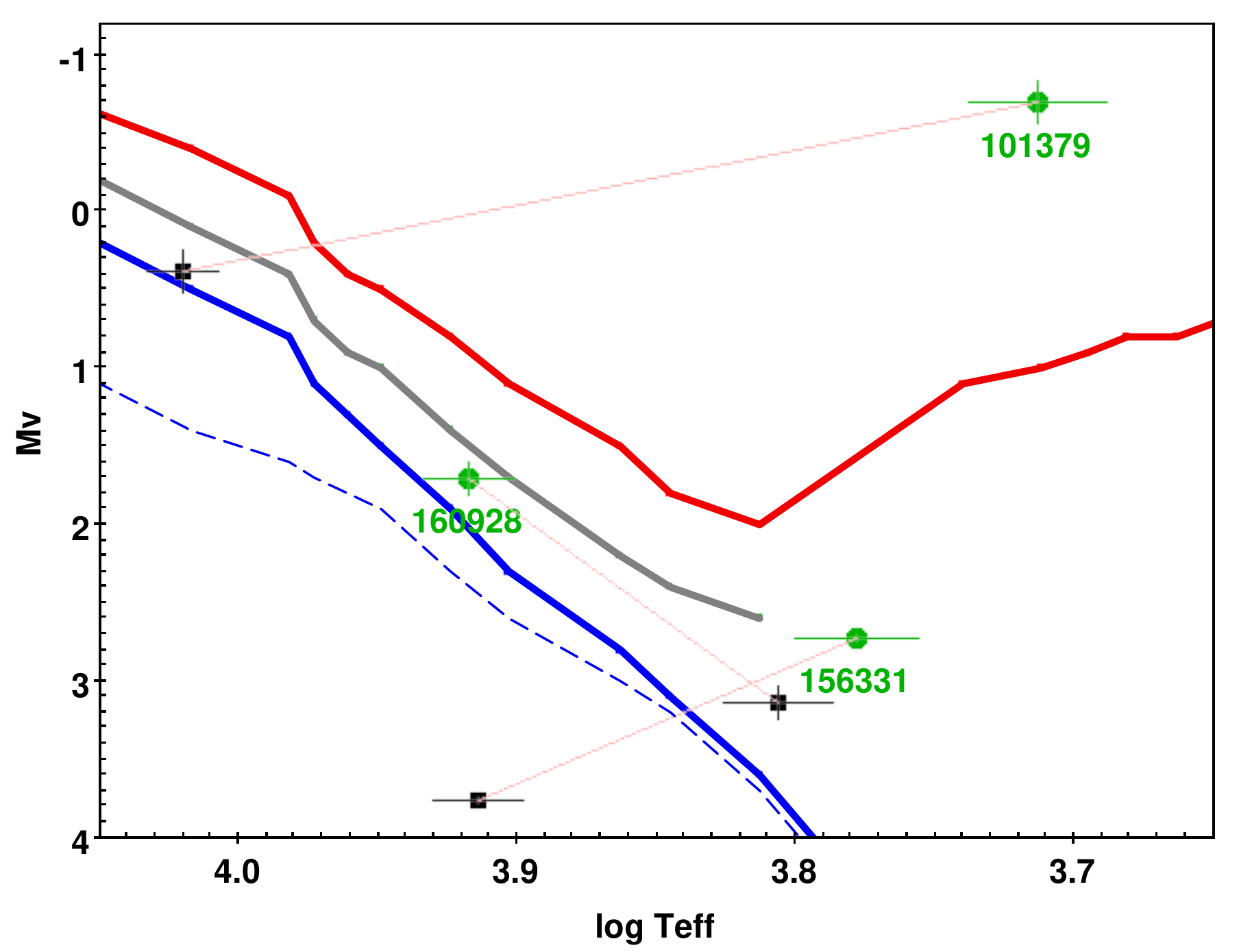}
    \includegraphics[width=8cm]{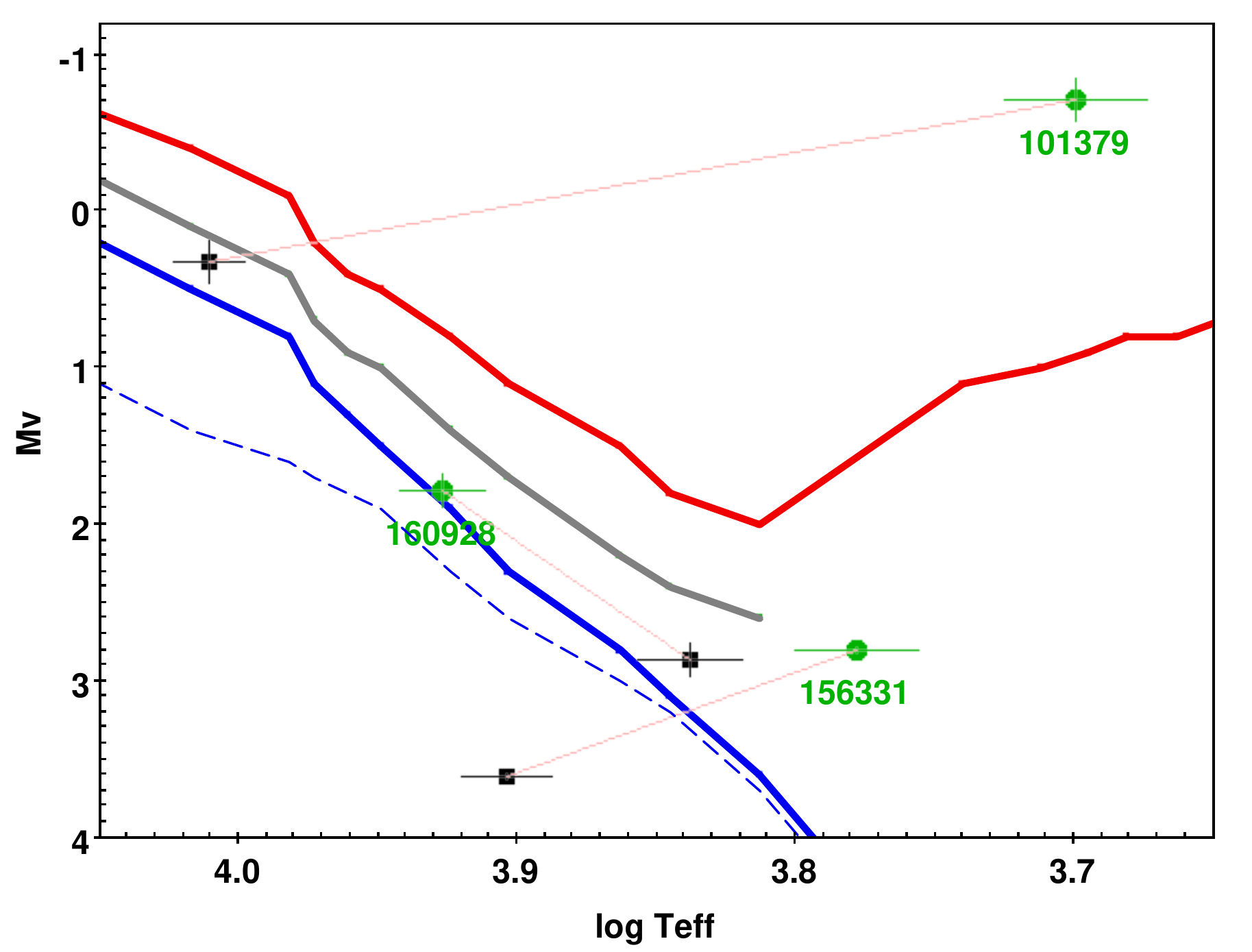}
    \caption{HD 101379, HD 156331 and HD 160928 systems on the HRD.
Solid curves represent main sequence, subgiant and
giant sequences (blue, grey and red, respectively),
and dashed blue curve represents ZAMS~\citep{1992msp..book.....S}.
Green circles and black squares represent primary (more luminous)
and secondary components of the binaries, respectively, with uncertainty bars.
Top and bottom panels: Coelho and Phoenix stellar models, respectively.}
    \label{fig:HRD}
\end{figure}



Analysis of totally reduced HRS spectra was done
with a dedicated software package 
{\tt fbs} \citep[Fitting Binary Stars;][]{2020RAA....20..119K}
developed by our team for the stellar spectra analysis
and used by us in the different studies
\citep{2019AstBu..74..183B,2019MNRAS.482.4408G,2020Ap&SS.365..169K,2021MNRAS.503.3856G}.
This software is fitted observed spectra with use of the library 
of high-resolution theoretical stellar spectra
and is designed to derive radial velocities
and stellar parameters (T$_{eff}$, log~g, sin~i and [Fe/H]) for each component
of a binary system. In clear cases of binary stars, the package uses
two model spectra with individual velocities and stellar parameters.
As an output software produces velocities and stellar parameters
for both components.
The current version of this software is using different stellar models
\citep{Coelho14,phoenix,tlusty}.

It was noted in \citet{2020Ap&SS.365..169K} that output errors of the {\tt fbs} are errors of fitting, 
which are often underestimate the real errors. Unfortunately, since each reduced \'echelle spectrum is
about hundred thousands points length, the way to find a global minimum of the function shown in
\citet{2020RAA....20..119K} is very time-consuming process.
For that reason, Monte-Carlo simulations or use of Markov chain Monte Carlo methods are not a real way
to estimate errors.
It is easier to estimate accuracy of the method comparing results of {\tt fbs}
with previously published data \citep[for example, see][]{2020RAA....20..119K} or 
to use each obtained spectrum of the same object as an independent one, model each with {\tt fbs} 
and study the output result as the statistical sample \citep{2020Ap&SS.365..169K}.

The results of our {\tt fbs} modeling for obtained spectra are presented in Table~\ref{tab:stars}
and shown in Fig.~\ref{fig:spectrum}. Since three spectra were obtained for HD\,101379,
each spectrum was analyzed with {\tt fbs} independently and presented parameters and their
errors are average values for this star. We also repeated the same analysis with both \citet{Coelho14} 
(\textsc{Coelho} in Table~\ref{tab:stars}) and 
\citep{phoenix} stellar models (\textsc{Phoenix} in Table~\ref{tab:stars}),
convolving them to match the HRS MR instrumental resolution. 
These results are also presented in Table~\ref{tab:stars}. 
Finally, after comparison these results we use errors for each parameter in thsi work
as they shown in the last row of Table~\ref{tab:stars}.

\section{Parameters of the studied systems}
\label{txt:results}

The parameters of three systems observed with SALT are shown in Fig.~\ref{fig:HRD}
and presented in Tables~\ref{tab:pairs} and~\ref{tab:stars}.
Absolute magnitudes of the components $M_V$ are calculated from parallax and visual brightness
(see Table~\ref{tab:pairs}) and from weight in V band and interstellar reddening $\rm E(B-V)$ values
(see Table~\ref{tab:stars}).


We can see from Fig.~\ref{fig:HRD} that more massive (and more luminous)
components of HD~101379 and HD~160928 look more evolved than the less massive ones.
It should be added that according to MSC~\citep{2018ApJS..235....6T},
HD~101379 (= WDS~11395-6524 = HIP~56862) is in a fact a quadruple system,
where HD~101379~A is a spectroscopic binary SB1, and HD~101379~B is an eclipsing binary.

Contrary, the secondary (less massive) component of HD~156331 is more evolved than
the primary (see Fig.~\ref{fig:HRD}),
and, consequently, it is a good candidate to the wide binary
with non-coeval components.
It is an indication of the non-coevality
of the components and, consequently, we can assume that this system
could be formed by a capture.

In addition, another assessment can be made.
HD~156331 was included in our list of pairs with probably non-coeval components~\citep{2020INASR...5..341M}
because the spectral classification of the components, F8III+B9V, given in WDS~\citep{2001AJ....122.3466M},
corresponds to masses of 1.76 and 2.58 (hereafter in solar mass), respectively, and hence
the less massive component looks more evolved.
The values of T$_{eff}$, log~g obtained in this work for this system
(see Table~\ref{tab:stars}, Phoenix library)
rather indicate the spectral types F5III+A7IV, which corresponds to masses of
1.51 and 1.82. The difference in masses has decreased, but the situation has not changed qualitatively.
Here we use the scales SpT -- T$_{eff}$ -- log~g -- mass from~\citep{1992msp..book.....S},
the typical mass error value when estimating it from the spectral type is 0.1.

However, the assumption that HD 156331 B has LC=IV or V
contradicts with its position on the HRD (see Fig.~\ref{fig:HRD}).
The star is located well below ZAMS. According to WDS~\citep{2001AJ....122.3466M},
the components of HD 156331 have the same brightness in the V-band
(which is contrary to our Weight values in Table~\ref{tab:stars}), but even under this assumption
absolute magnitude of HD~156331~B
``rises'' only to a value of $M_V$= 3.13 mag and the star remains under ZAMS.

In his detailed study of several fast visual binaries, based on speckle interferometry,
\cite{2017AJ....154..110T}, in particular, found orbital elements and calculated
dynamical parallax for HD~156331 ($\varpi_{dyn}$ = 14.2 mas), which turned out to be different
from the Hipparcos value ($\varpi_{HIP}$ = 16.9 mas). The recent Gaia observations rather confirm
the latter value ($\varpi_{Gaia}$ = 16.7 mas, see Table~\ref{tab:pairs} and
note that the parallax is given with a fairly high accuracy, 3\%).
But even relying on dynamical parallax will only make the components brighter by about 0.3 mag,
and HD~156331~B will still remain under ZAMS in Fig.~\ref{fig:HRD}.

Our values of HD 156331 components' radial velocities (see Table~\ref{tab:stars})
are quite consistent with RV difference of $\sim30$ km s$^{-1}$ near the periastron, predicted
by~\cite{2017AJ....154..110T}. Contrary, components' magnitude defference found
by~\cite{2017AJ....154..110T} corresponds to WDS values and not to values, found in this
paper and given in Table~\ref{tab:stars}.

Finally, it should be pointed out that the formal errors derived from the separation of the spectra
of the parameters of the components significantly underestimate the real ones.
Moreover, the output parameters can be correlated (for example, $\log g$ and $\rm V \sin i$),
so the analysis of the position of the components on the HRD must take into account
the errors of the total brightness, and of the magnitude difference, as well as errors
and a possible correlation of the obtained temperatures.

So the HD 156331 system requires further observation and analysis.

\section{Conclusion}
\label{txt:summary}

We have studied several stars from our preliminary list of candidates
to wide pairs with non-coeval components, and we have
found that the less massive component of HD~156331 is probably more
evolved than the more massive. 
We can assume that this system could be formed by a capture.
To prove the non-coevality one needs a detailed investigation of this
and other candidates.


\section{Acknowledgments}

{\akn We are grateful to our anonymous reviewer whose constructive comments greatly helped us to improve the paper.
All spectral observations reported in this paper were obtained with
    the Southern African Large Telescope (SALT) under program 2020-1-MLT-002 (PI: Alexei Kniazev).}
The work was partly supported by the Russian Foundation for Basic Researches (project 19-07-01198).
A.\,K. acknowledges support from the National Research Foundation (NRF) of South Africa. 
This research has made use of NASA's Astrophysics Data System,
of the SIMBAD database, operated at CDS (Strasbourg, France),
of TOPCAT, an interactive graphical viewer and editor for tabular data \citep{2005ASPC..347...29T}.
The acknowledgements were compiled using the Astronomy Acknowledgement Generator. 

\bibliographystyle{raa}
\bibliography{noncoev}

\end{document}